\documentclass[aps,pre,twocolumn,groupedaddress]{revtex4-2}
\usepackage{amsmath,amssymb}
\usepackage{slashed}
\usepackage{graphicx}
\graphicspath{ {./figures/} }
\usepackage{hyperref}
\usepackage{cleveref}
\usepackage{verbatim}
\usepackage{verbatim}

\usepackage{float}
\usepackage{subfig}

\usepackage[dvipsnames]{xcolor}

\newcommand{\dd}{\text{d}}
\renewcommand{\vec}[1]{\boldsymbol{#1}} 

\begin{document}
\title{The non-equilibrium solvent response force:\\
What happens if you push a Brownian particle}

\author{Fabian Koch}
\email[]{fabian.glatzel@physik.uni-freiburg.de}
\author{Jona Erle}
\author{Tanja Schilling}
\email[]{tanja.schiling@physik.uni-freiburg.de}
\affiliation{Institut f\"ur Physik, Albert-Ludwigs-Universit\"at Freiburg,\\ Hermann-Herder-Stra\ss e 3, 79104 Freiburg im Breisgau, Germany}

\date{\today}

\begin{abstract}
In this letter we discuss how to add forces to the Langevin equation.
We derive the exact generalized Langevin equation for the dynamics of one particle subject to an external force embedded in a system of many interacting particles. The external force may depend on time and/or on the phase-space coordinates of the system.  We construct a projection operator such that the drift coefficient, the memory kernel, and the fluctuating force  of the Generalized Langevin equation are the same as for the system without external driving. We show that the external force then enters the generalized Langevin equation additively. In addition we obtain one term which, to our knowledge, has up to now been overlooked. We analyze this additional term for an exemplary system.
\end{abstract} 
\maketitle

Langevin's description of the motion of a Brownian particle is a typical example of a coarse-grained model \cite{snook2006}. When modelling Brownian motion, the effect of the solvent degrees of freedom is not computed explicitly but replaced by two effective, coarse-grained quantities, the friction coefficient $\gamma$ and the random force $\vec{f}(t)$. The  velocity of a Brownian particle of mass $m$  then obeys the equation
\begin{equation}
\label{eq:SimpleLangevin}
     m\frac{\dd \vec{v}(t)}{\dd t} = - \gamma \vec{v}(t) + \vec{f}(t).
\end{equation}
The components of $\vec{f}(t)$ are related to $\gamma$ by the fluctuation dissipation theorem (FDT), $\langle f_i(t) f_j(s)\rangle = 2\gamma k_B T \delta_{i,j}\delta(t-s)$, where $k_B$ is Boltzmann's constant and $T$ is the temperature.

If the Brownian particle is subjected to an external force $\vec{F}_{\rm ext}$, this force often is simply added to eq.~\ref{eq:SimpleLangevin}. This is common practice when molecular dynamics (MD) simulations are carried out with a Langevin thermostat \cite{allen2007}, i.e.~it is done in a large number of MD simulation and Brownian Dynamics (BD) simulation studies of colloidal systems, biomolecular systems and polymeric systems. To our knowledge there is no derivation in the literature, which proves that this widely accepted procedure is correct -- or at least in which limits it might hold. 

Two seminal pieces of work are often cited to justify this approach: the article by Kubo on the fluctuation dissipation theorem \cite{kubo1966fluctuation} and the article by Zwanzig on the non-linear generalized Langevin equation \cite{zwanzig1973}. However, in neither of these articles such a statement is derived. Kubo explicitly wrote that he only considered linear effects in the external force, and Zwanzig considered particles in a bath, i.e.~the case in which the degrees of freedom which are integrated out do not interact with each other.
Other derivations like the one by Batchelor \cite{batchelor_1983} or the one by Hauge and Martin-L\"of \cite{hauge1973fluctuating} already start out from hydrodynamics, thus, they contain assumptions about local thermal averages and do not give a justification based on the full microscopic description.

In contrast, there is a piece of work in the literature that proves that forces cannot simply be added: a very readable article from 1972 by Kim and Oppenheim \cite{kim1972molecular}, which unfortunately has hardly ever been cited. In this article, the method famously used by Mazur and Oppenheim to derive the time-local Langevin equation as the Brownian limit of the generalized Langevin equation by an expansion in the ratio of the solvent particles' mass to the Brownian particles' mass \cite{mazur1970molecular} is extended to the case of an applied external force. One result of the derivation by Kim and Oppenheim is that the FDT does not hold if the dynamics are non-Markovian and there is an external force.

In this letter we recall why forces cannot simply be added on the level of the Langevin equation. Then we introduce a version of the projection operator method \cite{grabert2006projection} different from the one used in ref.~\cite{kim1972molecular} which allows to recover all terms in eq.~\ref{eq:SimpleLangevin} including the FDT and which does produce an additive force term, however, this comes at the cost of one additional term in the resulting Langevin equation. We then discuss the impact of this additional term on the resulting effective dynamics.

We start by briefly recalling Mori's work \cite{Mori_Transport_1965}: We consider the Hamiltonian equations of motion of $N$ interacting particles and integrate out all degrees of freedom apart from one component of the momentum of one particle. Under stationary conditions, the linear projection operator by Mori can be used for this task  and one obtains the linear generalized Langevin equation
\begin{align}
\label{eq:sLGLE}
    \frac{\dd p_z(t;\Gamma)}{\dd t} &= - \int\limits_0^t\dd\tau K(t-\tau) p_z(\tau;\Gamma) + \eta(t;\Gamma).
\end{align}
Here $p_z(t;\Gamma)$ denotes the momentum in $z$ direction at time $t$ given that the entire system was initialized at the phase-space point $\Gamma$ at time zero. $K(\tau)$ is the so-called memory kernel and $\eta(t;\Gamma)$ is the (deterministic) fluctuating force. The memory kernel and fluctuating force fulfill a fluctuation-kernel theorem
\begin{align}
    \left\langle \eta(t;\Gamma)\eta(t';\Gamma)\right\rangle &= K(t-t')\left\langle p_z^2\right\rangle.
\end{align}
(This type of equation is usually called a second fluctuation-dissipation theorem. However, as the structure holds generally for linear versions of the generalized Langevin equation irrespective of whether the kernel can be related to dissipation, we prefer to call it a fluctuation-kernel theorem (FKT).)

Now we address the following question: If one considers the same system but with an additional
external force acting on the particle of interest, could the motion be described
by an equation of the form
\begin{multline}
\label{eq:naiveGLE}
    \frac{\dd p_z(t;\Gamma)}{\dd t} = - \int\limits_0^t\dd\tau K(t-\tau) p_z(\tau;\Gamma)\\
    + \eta(t;\Gamma) + F_\text{ext}(t;\Gamma) \quad ?
\end{multline}
In the case of time-scale separation between $p_z$ and the other degrees of freedom, eq.~\ref{eq:SimpleLangevin} with an additive force would follow from \cref{eq:naiveGLE}. However, Mori's derivation only holds under stationary conditions, and it is not obvious that the effect of the external force can simply be added on the level of the coarse-grained description.

We first tackle the problem in more general terms and then treat \cref{eq:naiveGLE} as a special case. Given a system which is described by two Liouvillians, a time-independent Liouvillian $\mathcal{L}_0$ for the interactions within the system and $\mathcal{L}_1(t)$ for the time-dependent external force, the time evolution of an arbitrary observable $A(\Gamma)$ is given by
\begin{align}\label{eq:eom_step1}
	\frac{\dd A(t;\Gamma)}{\dd t} &= \mathcal{U}(t,t_0)\mathcal{L}_\text{tot}(t)A(\Gamma),
\end{align}
where $\mathcal{L}_\text{tot}(t)=\mathcal{L}_0+\mathcal{L}_1(t)$. Here, $\mathcal{U}(t,t_0)$ is the time-evolution operator that can be expressed as a negatively time-ordered exponential
\begin{align}
    \mathcal{U}(t,t_0) &=\exp_-\left(\int\limits_{t_0}^t\dd\tau \mathcal{L}_\text{tot}(\tau)\right)
\end{align}
and $\Gamma$ denotes the initial point in phase space at time $t_0$. 
Introducing a projection operator $\mathcal{P}$ and its orthogonal complement $\mathcal{Q}:=1-\mathcal{P}$, \cref{eq:eom_step1} can be rewritten as
\begin{align}
    \frac{\dd A(t;\Gamma)}{\dd t} &= \mathcal{U}(t,t_0)\mathcal{P}\mathcal{L}_0A(\Gamma) + \mathcal{U}(t,t_0)\mathcal{Q}\mathcal{L}_0A(\Gamma)\nonumber\\
    &\phantom{=}+ \mathcal{U}(t,t_0)\mathcal{L}_1(t)A(\Gamma).\label{eq:eom_step2}
\end{align}

We now rewrite second term on the right-hand side using a modified Dyson decomposition. To this end, we define
\begin{align}
     \mathcal{Z}(t,t_0) &:= \mathcal{U}(t,t_0)\mathcal{Q}
\end{align}
and calculate the time-derivative of $\mathcal{Z}(t,t_0)$ in order to obtain the differential equations
\begin{align}
     \!\!\frac{\partial Z(t,t_0)}{\partial t} &= \mathcal{U}(t,t_0)\mathcal{L}_\text{tot}(t)\mathcal{Q}\\
     &= \mathcal{Z}(t,t_0)\mathcal{L}_\text{tot}(t)\mathcal{Q}+\mathcal{U}(t,t_0)\mathcal{P}\mathcal{L}_\text{tot}(t)\mathcal{Q}\label{eq:standardDE}\\
     &=\mathcal{Z}(t,t_0)\mathcal{L}_0\mathcal{Q}+\mathcal{U}(t,t_0)(\mathcal{P}\mathcal{L}_0+\mathcal{L}_1(t))\mathcal{Q}.\label{eq:newDE}
\end{align}
In more general cases where the time-dependent part of the Hamiltonian/Liouvillian cannot be separated from the time-independent part as easily as here, \cref{eq:standardDE} is usually taken as the starting point to derive a generalized Langevin equation/Dyson decomposition \cite{doi:10.1063/1.5090450,PhysRevE.99.062118}. However, as the homogeneous part of the differential equation,
\begin{align}
     \frac{\partial Z_\text{hom}(t,t_0)}{\partial t} &= \mathcal{Z}_\text{hom}(t,t_0)\mathcal{L}_\text{tot}(t)\mathcal{Q},
\end{align}
already contains the external force through $\mathcal{L}_\textbf{tot}(t)$, it is not suitable to derive a generalized Langevin equation that has the same memory kernel as the stationary process as well as a fluctuation-kernel theorem. Instead we use \cref{eq:newDE}, where the $\mathcal{L}_1(t)$ contribution is shifted from the homogeneous part of the differential equation to the inhomogeneity. In this case the solution to the homogeneous differential equation reads
\begin{align}
     \mathcal{G}(t-t_0) &:= \exp\left((t-t_0)\mathcal{L}_0\mathcal{Q}\right),
\end{align}
which can be used to find the special solution to the inhomogeneous equation
\begin{align}
     \mathcal{Z}(t,t_0) &= \mathcal{U}(t',t_0)\mathcal{Q}\mathcal{G}(t-t')\nonumber\\
     &\phantom{=}+\int\limits_{t'}^t\dd\tau \mathcal{U}(\tau,t_0)(\mathcal{P}\mathcal{L}_0+\mathcal{L}_1(\tau))\mathcal{Q}\mathcal{G}(t-\tau).
\end{align}
It is easy to check that this expression solves the differential equation \cref{eq:newDE} together with the boundary condition $\mathcal{Z}(t',t_0) = \mathcal{U}(t',t_0)\mathcal{Q}$ (cf. ref.~\cite{evans1990statistical}) and, hence, one can substitute $\mathcal{U}(t,0)\mathcal{Q}$ in \cref{eq:eom_step2} with the new expression. One obtains the following equation of motion:
\begin{widetext}
    \begin{align}
        \frac{\dd A(t;\Gamma)}{\dd t} &= \mathcal{U}(t,t_0)\mathcal{P}\mathcal{L}_0A(\Gamma) +\int\limits_{t'}^t\dd\tau\, \mathcal{U}(\tau,t_0)\mathcal{P}\mathcal{L}_0\mathcal{Q}\mathcal{G}(t-\tau)\mathcal{L}_0A(\Gamma)+\mathcal{U}(t',t_0)\mathcal{Q}\mathcal{G}(t-t')\mathcal{L}_0A(\Gamma)\nonumber\\
	   &\phantom{=} + \mathcal{U}(t,t_0)\mathcal{L}_1(t)A(\Gamma)+\int\limits_{t'}^t\dd\tau\, \mathcal{U}(\tau,t_0)\mathcal{L}_1(\tau)\mathcal{Q}\mathcal{G}(t-\tau)\mathcal{L}_0A(\Gamma).\label{eq:eom_step3}
\end{align}
\end{widetext}
To make use of this expression one needs to find a suitable projection operator. We use a linear Mori projection operator of the form
\begin{align}
	\mathcal{P}X(\Gamma) &:= \frac{(X,A)}{(A,A)}A(\Gamma),
\end{align}
with
\begin{align}
	(X,Y):=\int\dd\Gamma\,\rho_0(\Gamma)X(\Gamma)Y(\Gamma).
\end{align}
Here, $\rho_0(\Gamma)$ is the phase-space distribution at time $t_0$, which needs to be stationary under the dynamics $\mathcal{L}_0$, but which is not stationary under 
$\mathcal{L}_\text{tot}(t)$. Using this projection operator, \cref{eq:eom_step3} becomes
\begin{align}
    \frac{\dd A(t;\Gamma)}{\dd t} &= c A(t;\Gamma) - \int\limits_{t'}^t\dd\tau\,K(t-\tau)A(\tau;\Gamma)+\eta(t,t';\Gamma)\nonumber\\
	&\phantom{=}+ \mathcal{U}(t,t_0)\mathcal{L}_1(t)A(\Gamma) +F_\text{NER}(t,t';\Gamma),\label{eq:generalGLE}
\end{align}
with
\begin{align}
	c &:=\frac{(\mathcal{L}_0A,A)}{(A,A)},\\
	K(t) &:= -\frac{(\mathcal{L}_0\mathcal{Q}\mathcal{G}(t)\mathcal{L}_0A,A)}{(A,A)},\\
	\eta(t,t';\Gamma) &:= \mathcal{U}(t',t_0)\mathcal{Q}\mathcal{G}(t-t')\mathcal{L}_0A(\Gamma),\\
    F_\text{NER}(t,t';\Gamma) &:=\int\limits_{t'}^t\dd\tau\, \mathcal{U}(\tau,t_0)\mathcal{L}_1(\tau)\mathcal{Q}\mathcal{G}(t-\tau)\mathcal{L}_0A(\Gamma).
\end{align}
Note that the drift $c$ does not contain $\mathcal{L}_1(t)$. 
If we set $t'=t_0$ (and hence $\mathcal{U}(t',t_0)=1$), the memory kernel $K(t)$ and the fluctuating force $\eta(t,t_0;\Gamma)$ do not depend on $\mathcal{L}_1(t)$, either. Thus they are defined exactly as in the case of stationary dynamics under $\mathcal{L}_0$. The memory kernel can then be written in the form
\begin{align}
	K(t) &= \frac{(\mathcal{Q}\mathcal{G}(t)\mathcal{L}_0A,\mathcal{Q}\mathcal{L}_0A)}{(A,A)}\\
	&=\frac{\langle\eta(t,0;\Gamma),\eta(0,0;\Gamma)\rangle}{(A,A)},
\end{align}
i.e., the FKT still holds true. 

If now the observable of interest is a component of the momentum of one particle, $A(\Gamma)=p_z$, and the Liouvillian $\mathcal{L}_1(t)$ is an external force which acts on this degree of freedom, e.g.~$\mathcal{L}_1(t) = F_\text{ext}(t)\partial_{p_z}$ or $\mathcal{L}_1(t)=-\frac{\partial V_\text{ext}(z)}{\partial z}\partial_{p_z}$, \cref{eq:generalGLE} turns into
\begin{align}
    \frac{\dd p_z(t;\Gamma)}{\dd t} &= - \int\limits_0^t\dd\tau\,K(t-\tau)p_z(\tau;\Gamma)+\eta(t,0;\Gamma)\nonumber\\
    &\phantom{=}+ F_\text{ext}(t,z(t)) + F_\text{NER}(t,0;\Gamma),
    \label{eq:generalGLE_p}
\end{align}
where $K$ and $\eta$ only contain the propagator $\mathcal{L}_0$, while the variable $p_z$ is propagated by $\mathcal{L}_0 + \mathcal{L}_1$.
We have thus succeeded in constructing an exact generalized Langevin equation for a particle subjected to an external force, which contains the same memory kernel and fluctuating force as the generalized Langevin equation for the particle without the external force, \cref{eq:sLGLE}. However, compared to \cref{eq:naiveGLE} there is one additional term, the non-equilibrium response force $F_\text{NER}(t,0;\Gamma)$. This force encodes the non-equilibrium response of the solvent to the driven particle. In general, it will depend on time and on the history of the process, and, as we will show below, it cannot be absorbed in the fluctuating force. 

To determine $F_\text{NER}(t,0;\Gamma)$ numerically, we make use of the fact that the functions $K(t)$ and  $\eta(t,0;\Gamma)$ in the stationary GLE, \cref{eq:sLGLE}, and in \cref{eq:generalGLE_p} are identical. We run equilibrium simulations, "measure" $p_z$ and compute the memory kernel by means of the method described in ref.~\cite{10.1002/adts.202000197}. The fluctuating forces for each trajectory then follow from \cref{eq:sLGLE}. Then we run simulations with an external force for the exact same starting configurations and calculate $F_\text{NER}(t,0;\Gamma)$ via \cref{eq:generalGLE_p}. 

\begin{figure}
    \centering
    \includegraphics[width=\linewidth]{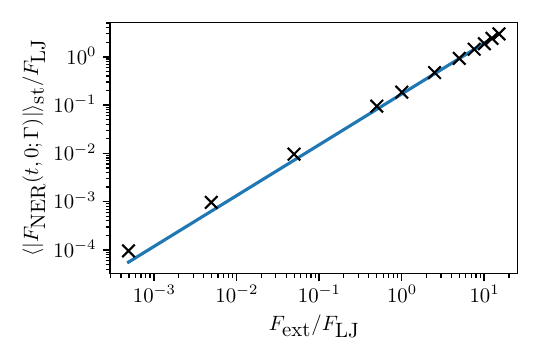}
    \caption{The non-equilibrium ensemble average of the magnitude of the NER force, $\langle|F_\text{NER}(t,0;\Gamma)|\rangle_\text{st}$, as a function of the external force. The blue line is a fit of the form $y=a F_\text{ext}^b$ where $b\approx1$.}
    \label{fig:force_scaling}
\end{figure}

\begin{figure}
    \centering
    \includegraphics[width=\linewidth]{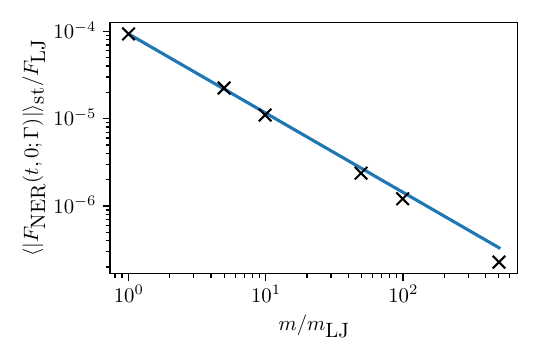}
    \caption{The non-equilibrium ensemble average of the magnitude of the NER force, $\langle|F_\text{NER}(t,0;\Gamma)|\rangle_\text{st}$, as a function of the mass ratio. The blue line is a fit of the form $y=a m^b$ where $b\approx-1$.}
    \label{fig:mass_scaling}
\end{figure}

The simulations contain $1024$ particles interacting via a Lennard-Jones potential. We use Lennard-Jones units and define a Lennard-Jones force $F_\text{LJ}:=m_\text{LJ}\sigma_\text{LJ}/\tau_\text{LJ}^2$.
In this letter, we restrict the discussion to the case of a constant force applied for times $t>0$ acting in z-direction $F_\text{ext}(t,z(t)) = F_\text{ext}\theta(t)$. In all our simulations, we found $F_\text{NER}(t,0;\Gamma)$ to be non-zero.
In order to determine how the term vanishes with $F_\text{ext}\to 0$, we calculate the ensemble average of the magnitude $\langle|F_\text{NER}(t,0;\Gamma)| \rangle_\text{st}$, where the brackets indicate the average over the non-equilibrium ensemble of trajectories initialized according to a canonical distribution with respect to $\mathcal{L}_0$. Further, a time-average was taken over a short time interval (0.2 $\tau_{\text{LJ}}$) in order to smoothen the data. \Cref{fig:force_scaling,fig:mass_scaling} show that this average vanishes linearly with the external force as well as with the mass ratio.

\begin{figure}
    \centering
    \includegraphics[width=\linewidth]{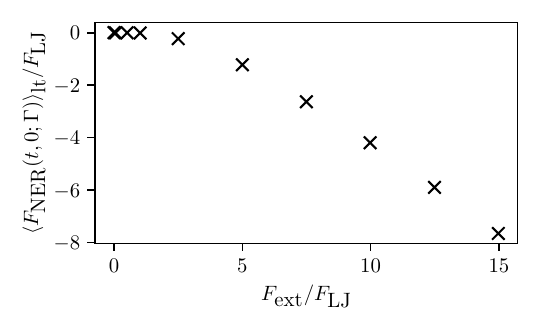}
    \caption{The long-time ensemble average of the NER force, $\langle F_\text{NER}(t,0;\Gamma)\rangle_\text{lt}$, as a function of the external force.}
    \label{fig:crossover}
\end{figure}

\begin{figure}
    \centering
    \includegraphics[width=\linewidth]{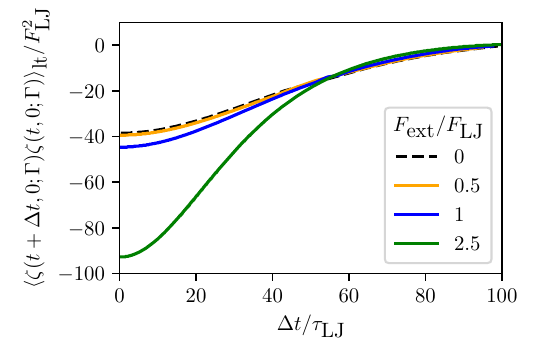}
    \caption{The plot shows the auto-correlation function function of $\zeta_\perp(t,0;\Gamma)$ vs. the time difference $\Delta t$ for different constant external forces.}
    \label{fig:fkr}
\end{figure}

If there is a force missing in the Langevin equation, why do BD simulations and MD simulations with a Langevin thermostat work? After all, these methods are widely used for systems under external forces (see e.g.~BD simulations of colloidal particles in optical traps \cite{volpe2013}, Langevin MD simulations of proteins pulled through pores \cite{luo2007} or targeted MD simulations in the context of biomolecular modelling \cite{wolf2018}). One conjecture could be that the stochastic interpretation of the fluctuating force absorbs the effect, i.e.~when the deterministic fluctuating force $\eta$ is replaced by a noise, perhaps the resulting dynamics is the same as if $\zeta(t,0;\Gamma):=\eta(t,0;\Gamma) +F_\text{NER}(t,0;\Gamma)$ were replaced by the same noise. 
For this to hold, the sum $\zeta(t,0;\Gamma)$ must exhibit the same statistical properties as $\eta(t,0;\Gamma)$ itself. We now check if this conjecture holds, i.e.~if $\zeta(t,0;\Gamma)$ fulfills the FKT and if $\langle\zeta(t,0;\Gamma)\rangle = 0$. Since the ensemble average is linear and $\langle \eta(t,0;\Gamma)\rangle=0$, it suffices to check if $\langle F_\text{NER}(t,0;\Gamma) \rangle$ vanishes. For all external forces considered here, we see that the ensemble average of the NER force converges to a constant. \Cref{fig:crossover} shows this limit, calculated by taking the time average of $\langle F_\text{NER}(t,0;\Gamma) \rangle$ for large times ($3\leq t/\tau_\text{LJ}\leq25$). There is a crossover point when the external force has the same magnitude as the fluctuating forces, beyond which the magnitude of $\langle F_\text{NER}(t,0;\Gamma) \rangle$ increases linearly. Since the external force was applied in positive $z$-direction, this plot also shows, that on average, the NER force counteracts the external force, making it relevant for the calculation of friction. 

One can observe a similar trend by testing the FKT for $\zeta(t,0;\Gamma)$, cf. \cref{fig:fkr}. The correlation functions of $\zeta(t,0;\Gamma)$ are calculated in the long-time limit where the ensemble average of $\zeta(t,0;\Gamma)$ has reached its plateau. For small external forces the FKT seems to be fullfilled. However, already at $F_\text{ext}=F_\text{LJ}$ a clear deviation can be seen. Thus we conclude that the conjecture does not hold. The NER force is not absorbed in the noise.

Note that in case the external force varies with time, $F_\text{NER}$ does not necessarily reach a stationary limit and the dependence of $F_\text{NER}$ on the mass ratio and on the magnitude of the force will be more complex. Recent work by Espa\~nol on the GENERIC framework under external forcing implies that even in the Markovian limit the dynamics is non-trivial \cite{espanol2023statistical}. 

We conclude that the application of an external driving force produces an additional force on the level of the Langevin equation. This additional force, which encodes the non-equilibrium response of the solvent, is in general non-local in time. Its statistics is different from the statistics of the fluctuating force. 
It is probably possible to measure this force in single molecule pulling experiments or in experiments on tracer particles driven through complex liquids following the same protocol that we applied in our simulations. Considering simulations, we stress that
the application of a Langevin thermostat to MD simulations of particles which are not subject to external forces is, of course, perfectly fine, as $F_\text{NER}$ does not appear there. However, the application to more complex systems is less trivial. Regarding simulations of interacting particles immersed in a solvent, the impact of the potential of mean force on the structure of the Langevin equation has recently been discussed elsewhere \cite{glatzel2022interplay,vroylandt2022position,ayaz2022generalized}. The statement made in this letter refers to systems under external driving such as e.g.\ particles in optical traps or biomolecules in constrained MD simulations. Here, if one wishes to obtain quantitative results, care is required. In general, it is misleading to interpret the Langevin equation as a Newtonian equation of motion.

\subsection*{Acknowledgement}
The authors acknowledge funding by the Deutsche Forschungsgemeinschaft (DFG, German Research Foundation) in Project No. 430195928. They also thank H.~Meyer, J.~Dzubiella, S.~Milster and C.~Widder for stimulating discussions.

\end{document}